\documentclass{elsart}
\usepackage{amssymb}

\begin{document}
 
\journal{Physics Letters A}
 
\begin{frontmatter}
 
\title{ Deterministic diffusion in one-dimensional maps -- Calculation
of diffusion constants by harmonic inversion of periodic orbit sums   }
\author{Kirsten Weibert, J\"org Main, and G\"unter Wunner}
\address{Institut f\"ur Theoretische Physik 1,
         Universit\"at Stuttgart, D-70550 Stuttgart, Germany}
\maketitle
 
\begin{abstract}
A method is proposed for the calculation of diffusion constants
for one-dimensional maps exhibiting deterministic diffusion.
The procedure is based on harmonic inversion and uses a known relation
between the diffusion constant and the periodic orbits of a map.
The method is tested on an example map for which results 
calculated by different other techniques are available for comparison.
\end{abstract}
 
\end{frontmatter}

\section{Introduction}
Deterministic diffusion plays an important r\^ole in a variety
of physical phenomena and applications, e.g. particle confinement
in magnetic fields \cite{Chi79}, conductivity in metals
\cite{Wag91}, etc.
A general theory of diffusion for systems with continuous time is
presented in \cite{Gas96}.
In systems with periodic phase-space structure, diffusion
may be induced by chaotic dynamics inside the elementary cells,
causing an irregular jumping between the cells.
The investigation of deterministic diffusion is an interesting task
both from the theoretical point of view, concerning the question
of how the typical stochastic properties arise from a purely deterministic
dynamics, as well as with regard to its various applications
in different physical contexts.

It has been observed that even the simplest model systems
-- one-dimensional maps -- which exhibit deterministic diffusion
are capable of explaining the behaviour of relevant physical
systems \cite{Art93,Eck93}.
For such maps, it has been demonstrated that
the diffusive properties are related
to the periodic orbits of the map
restricted to an elementary cell \cite{Art93,Eck93,Art91,Cvi01}.
In particular, the diffusion constant characterizing the 
diffusive spreading
can be obtained from the leading zero of a zeta function consisting
of a product over contributions from all periodic orbits of the reduced map.
Similar to the dynamical zeta functions which appear in semiclassical 
quantization problems, this product usually does not converge.
In former work \cite{Art93,Eck93,Art91}, 
the convergence problem was handled by the application
of cycle expansion techniques. However, cycle expansion 
strongly depends on the existence of a symbolic dynamics 
and the shadowing of long orbits by combinations of short ones.

On the other hand, it has been demonstrated  \cite{Mai97b,Mai99a}
that the convergence problems
of the zeta functions and corresponding response functions
arising in semiclassical quantization problems
can be circumvented by the application of 
harmonic inversion, which is a universal method in the sense that it
does not rely on a symbolic code or other special properties of the system.
The method has recently been extended from Hamiltonian systems
to maps \cite{Wei01a}.
In this paper we demonstrate that harmonic inversion can also be used
for the calculation of diffusion constants of one-dimensional maps.
In analogy with the procedure for semiclassical quantization, we
rewrite the zeta function as a response function, from which we construct
a periodic orbit signal including a finite set of periodic orbits.
The analysis of the signal by harmonic inversion yields the zeros of the
underlying zeta function, from which the diffusion constant is
calculated. We apply this technique to an example map for which
the periodic orbits can easily be calculated, and which has already
been examined with the help of
cycle expansion techniques \cite{Det97},
rendering a comparison with previous results possible.

We start by describing the general procedure.
We consider one-dimensional maps 
$\hat f: \mathbb R \to \mathbb R$
defined by an anti-symmetric function
$\hat f$,
\begin{equation}
\hat x_{n+1}=\hat f (\hat x_n)\; ,
\qquad \hat f(-\hat x)= -\hat f (\hat x)
\end{equation}
for points $\hat x$ inside the elementary cell $[-1/2,1/2)$, 
and a discrete translational symmetry
outside the elementary cell,
\begin{equation}\label{^f^x}
\hat f(\hat x + n) = \hat f(\hat x) + n\; , \qquad n\in {\mathbb Z}\; .
\end{equation}
From this map, a reduced map $f: [-1/2,1/2) \to [-1/2,1/2)$
can be constructed by restricting the dynamics
to the elementary cell via the definition
\begin{eqnarray}
x & = & \hat x- [\hat x+1/2] \label{x^x} \\
f(x)  & = &  \hat f(x)- [\hat f(x)+1/2]\; ,
\label{fx}
\end{eqnarray}
where $[\, \cdot\, ]$ denotes the integer part of the quantity in the brackets.
Each periodic orbit of the reduced map $f(x)$,
consisting of a sequence of fixed points $\{x_1,x_2,\dots,x_n\}$,
corresponds to an orbit of the full map $\hat f(\hat x)$ 
which returns to an equivalent point,
\begin{equation}
\hat f^n(x_j)=x_j+\sigma_p\; ,\qquad \sigma_p\in {\mathbb Z}\; .
\end{equation}
The diffusion constant associated with the map is determined by the
lattice translation $\sigma$ after $n$ applications of the map, 
averaged over initial
conditions in the elementary cell
\begin{equation}\label{defD}
D=\lim_{n\to\infty} {1\over 2n} \langle \sigma^2 \rangle_n \; .
\end{equation}

As was shown in Ref.~\cite{Art91}, 
for hyperbolic or almost hyperbolic maps,
the diffusion constant can be determined from the periodic orbits
of the reduced map (\ref{fx})
by considering the zeta function
\begin{equation}\label{zeta_func}
Z(z,\beta)=\prod_{m=0}^\infty \prod_p 
\left( 1 - {z^{n_p}\ e^{\sigma_p \beta}\over 
|\Lambda_p|\ \Lambda_p^m} \right)\; ,
\end{equation}
where the second product runs over all primitive periodic orbits
of the reduced map; 
$n_p$ and $\Lambda_p=\sum_{i=1}^nf'(x_i)$ are the topological length and 
the stability
of the orbit, respectively, 
$\sigma_p$  denotes the lattice translation of the corresponding orbit
of the full map, 
and $\beta$ is a free parameter.
According to Ref.~\cite{Art91},
the diffusion constant $D$ can 
be expressed in terms of the leading zero
$z_0(\beta)$ of the zeta function as a function of $\beta$,
\begin{equation}\label{d2z}
D\approx -{1\over 2}\, 
\left[{\partial^2\over \partial\beta^2}\  z_0(\beta)
\right]_{\beta=0}\; .
\end{equation}
Additionally,
from the symmetry of the map, it follows that 
$[\partial z_0/\partial\beta]_{\beta =0}=0$.
Using these relations, an expansion of $\ln z_0(\beta)$
for small $\beta$ yields
\begin{equation}\label{D-z}
\ln z_0(\beta) \approx - \beta^2 D\; ,
\end{equation}
which can directly be used to calculate the diffusion constant
from the leading zero for small $\beta$.

The problem that remains is how to actually calculate the leading
zero, since the expression
(\ref{zeta_func}) for the zeta function usually diverges.
In past works, this problem has been tackled by application of cycle expansion
techniques. 
As an alternative, we demonstrate in the following how the zeta function 
can be expressed in terms of a response function,
and how the leading zero can then be obtained by harmonic inversion
of a periodic orbit signal constructed from this response function.

\section{Harmonic inversion for maps}
We start by defining the quantity $w = -i\ \ln z$
and rewriting the zeta function (\ref{zeta_func}) as a function of $w$,
\begin{equation}\label{zeta_func2}
Z(w,\beta)=\prod_{m=0}^\infty \prod_p 
\left( 1 - {e^{i w n_p}\ e^{\sigma_p \beta}\over 
|\Lambda_p|\ \Lambda_p^m} \right)\; .
\end{equation}
The density of zeros of the zeta function (\ref{zeta_func2})
can be written in the form
\begin{equation}
\rho(w,\beta)=-{1\over \pi}\ {\rm Im}\ g(w,\beta)\; ,
\end{equation}
where the response function $g(w,\beta)$ is given by
\begin{equation}
g(w,\beta)=g_0(w,\beta)+{\partial \over \partial w}\,  \ln Z(w,\beta) \; 
\end{equation}
with $g_0(w,\beta)$ a smooth, slowly varying function of $w$. 
Using the definition
\begin{equation}
t_p^{(m)}=   {e^{i w n_p}\ e^{\sigma_p \beta}\over 
|\Lambda_p|\ \Lambda_p^m}\; ,
\end{equation}
we rewrite the second term of the response function as
\begin{eqnarray}
{\partial \over \partial w}\,  \ln Z(w,\beta)&=&
\sum_{m=0}^\infty \sum_p (-i n_p)\  {t_p^{(m)}\over 1-  t_p^{(m)}}\\
&=&
 \sum_{m=0}^\infty \sum_p \sum_{r=1}^\infty (-i n_p)
\left(t_p^{(m)}\right)^r\\
&=&
 \sum_{m=0}^\infty \sum_p \sum_{r=1}^\infty (-i n_p)\ 
{e^{i w rn_p}\ e^{r\sigma_p \beta}\over 
|\Lambda_p^r|\ \Lambda_p^{rm}} \\
&=&
 \sum_{\rm po}
(-i n_0)\ { e^{i w n_{\rm po}}\ e^{\sigma_{\rm po} \beta} 
\over |\Lambda_{\rm po}|}
 \sum_{m=0}^\infty \Lambda_{\rm po}^{-m}\; ,
\label{g3}
\end{eqnarray}
where the first sum in (\ref{g3})
now runs over {\it all} periodic orbits (p.o.) including multiple
traversals of primitive orbits, and $n_0$ denotes the topological length
of the underlying primitive orbit.
After carrying out the sum over $m$, the response function finally reads
\begin{equation}\label{gpo}
g(w,\beta) = g_0(w,\beta)+\sum_{\rm po}
(-i n_0)\ { e^{\sigma_{\rm po} \beta} \over |1-\Lambda_{\rm po}|}\ 
e^{iwn_{\rm po}} \; .
\end{equation}
On the other hand, the response function can be written in terms
of the zeros  $w_k$ of the zeta function (\ref{zeta_func2})
and their multiplicities $d_k$,
\begin{equation}\label{gex}
g(w,\beta)=\sum_k {d_k \over w - w_k(\beta)} \; .
\end{equation}

Like the zeta function (\ref{zeta_func2}), the expression (\ref{gpo})
for the response function will usually not converge in those regions
of the complex variable $w$, where the poles of $g(w,\beta)$ are located, 
and the zeros of the zeta function cannot directly be calculated from 
(\ref{gpo}).
Instead, we apply harmonic inversion to obtain an analytical continuation
\cite{Mai97b,Mai99a}.
The poles $w_k(\beta)$ of the analytical continuation of $g(w,\beta)$
in (\ref{gpo}) are the zeros of the analytical continuation of the 
zeta function $Z(w,\beta)$ in (\ref{zeta_func2}).
To apply harmonic inversion, we now construct a periodic orbit signal 
by Fourier transformation of the response function (\ref{gpo}),
\begin{equation}
C(s,\beta)= {1\over 2\pi} \int_{-\infty}^\infty
g(w,\beta)\ e^{-isw} dw  \: .
\end{equation}
Considering only the second, oscillating part of $g(w,\beta)$, this results in
\begin{equation}\label{cpo}
C(s,\beta)=
\sum_{\rm po} {-i n_0\ e^{\sigma_{\rm po} \beta} \over |1-\Lambda_{\rm po}|}\ 
\delta(s-n_{\rm po})\; .
\end{equation}
The corresponding signal resulting from Fourier transformation of 
Eq.~(\ref{gex}) reads
\begin{equation}\label{cex}
C(s,\beta)=-i\sum_k d_k\ e^{-isw_k(\beta)}\; ,
\end{equation}
which contains the zeros $w_k(\beta)$ of the zeta function (\ref{zeta_func2}) 
as frequencies.
The zeros $w_k(\beta)$ can now be determined
by adjusting the periodic orbit signal (\ref{cpo}) 
-- including all orbits up to a maximum topological length $n_{\rm max}$ -- 
to the form of the 
corresponding signal (\ref{cex})
by harmonic inversion.
According to Eq.~(\ref{D-z}) and with the relation $w=-i\ \ln z$, 
the diffusion constant can be obtained from the leading frequency
$w_0$ using the relation
\begin{equation}\label{D-w0}
D\approx {-i\over \beta^2}\ w_0(\beta)\; ,
\end{equation}
which for small $\beta$ should be independent of $\beta$.

\section{Results}
As an example we consider the following map, which has also been 
examined in Ref.~\cite{Det97}: For points $\hat x$ inside the interval
$[-1/2,1/2]$ the map is defined by
\begin{equation}\label{map}
\hat f(\hat x)= \hat x (1+ 2|2\hat x|^\alpha)\; ,
\end{equation}
where $\alpha$ is a free parameter with the restriction $\alpha >-1$.
Outside the elementary cell, the map is continued according
to Eq.~(\ref{^f^x}).
Using Eqs.~(\ref{x^x}) and (\ref{fx}), a corresponding reduced
map $f(x)$ can be constructed.

As was discussed in Ref.~\cite{Det97}, 
the periodic orbits of the reduced
map $f(x)$ can be described by a complete
symbolic dynamics with a three-letter alphabet.
For $-1 < \alpha \le 0$ the dynamics is completely hyperbolic.
The 0 orbit (corresponding to the fixed point $x=0$)
is infinitely unstable (i.e., $\Lambda_p=\infty$) 
and does not contribute to the response function (\ref{gpo}).
For $\alpha > 0$, the 0 orbit becomes marginally stable
with $\Lambda_p=1$, and the dynamics
becomes intermittent.
Furthermore, it was shown
in \cite{Det97}  that at $\alpha=1$ the system
undergoes a phase transition 
from normal to anomalous diffusion,
and the diffusion constant as defined by Eq.~(\ref{defD})
equals 0 for $\alpha\ge 1$.

For $\alpha > 0$,
the marginally stable 0 orbit would give a diverging contribution to the
zeta function (\ref{zeta_func}) as well as to the periodic orbit signal
(\ref{cpo}). 
As was discussed in Ref.~\cite{Cvi01}, it is appropriate to simply 
omit this orbit 
from the zeta function,
as was also done in Ref.~\cite{Det97}.
In Ref.~\cite{Det97}, the diffusion constant of the map 
(\ref{map}) as a function of $\alpha$ was determined 
using explicit cycle expansion expressions for the average 
in Eq.~(\ref{defD})
rather than making use of Eq.~(\ref{d2z}).
In the following, we will determine the diffusion constant
by harmonic inversion as outlined above.

\begin{figure}[p]
\vspace{15cm}
\includegraphics{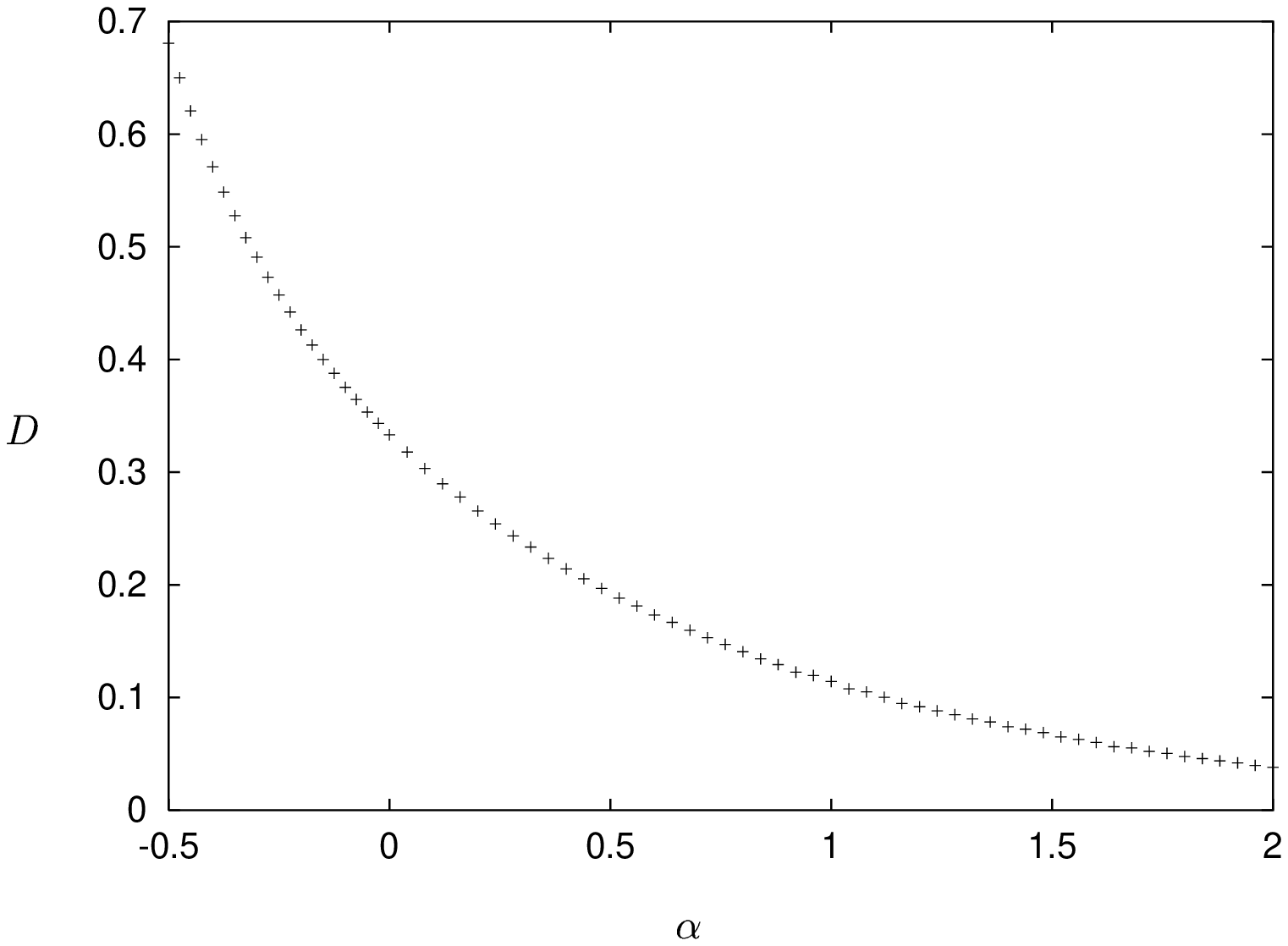}
\caption{Results for the diffusion constant of the map (\ref{map})
as a function of the parameter $\alpha$, calculated by harmonic inversion of
the periodic orbit signal (\ref{cpo}) including all periodic 
orbits up to topological length 13 (with the exception of the 0 orbit).
}
\label{fig1}
\end{figure}

We calculated the diffusion constant for different values of the
parameter $\alpha$ in the region $-0.5\le \alpha \le 2$ 
using the periodic orbits up to topological length $N=13$
(but omitting the 0 orbit).
For each value of $\alpha$, we performed calculations for different
small values of $\beta$ and checked the convergence of the
results for the diffusion constant for $\beta\to 0$.
The results for the diffusion constant are presented in Figure \ref{fig1}.
In the region  $\alpha\lesssim 0.5$,
our results are in excellent agreement with those from Ref.~\cite{Det97}
obtained with the periodic orbits up to topological length $N=10$.
For $\alpha> 0.5$, the different techniques applied
in Ref.~\cite{Det97} yielded quantitatively different values for
the diffusion constant; here, our results show a good qualitative agreement
with the different cycle expansion results.
This is even true when we used the reduced set of orbits up to topological 
length $N=10$ in our calculations: The function $D(\alpha)$ in Fig.~\ref{fig1}
becomes slightly more flat, but still shows a good qualitative 
agreement with the results of Ref.~\cite{Det97}.
As in \cite{Det97}, it was not possible to reproduce the 
theoretical value $D=0$ for $\alpha\ge 1$, but one only obtains an
asymptotical convergence to this value.

\begin{figure}[p]
\vspace{19.5cm}
\includegraphics{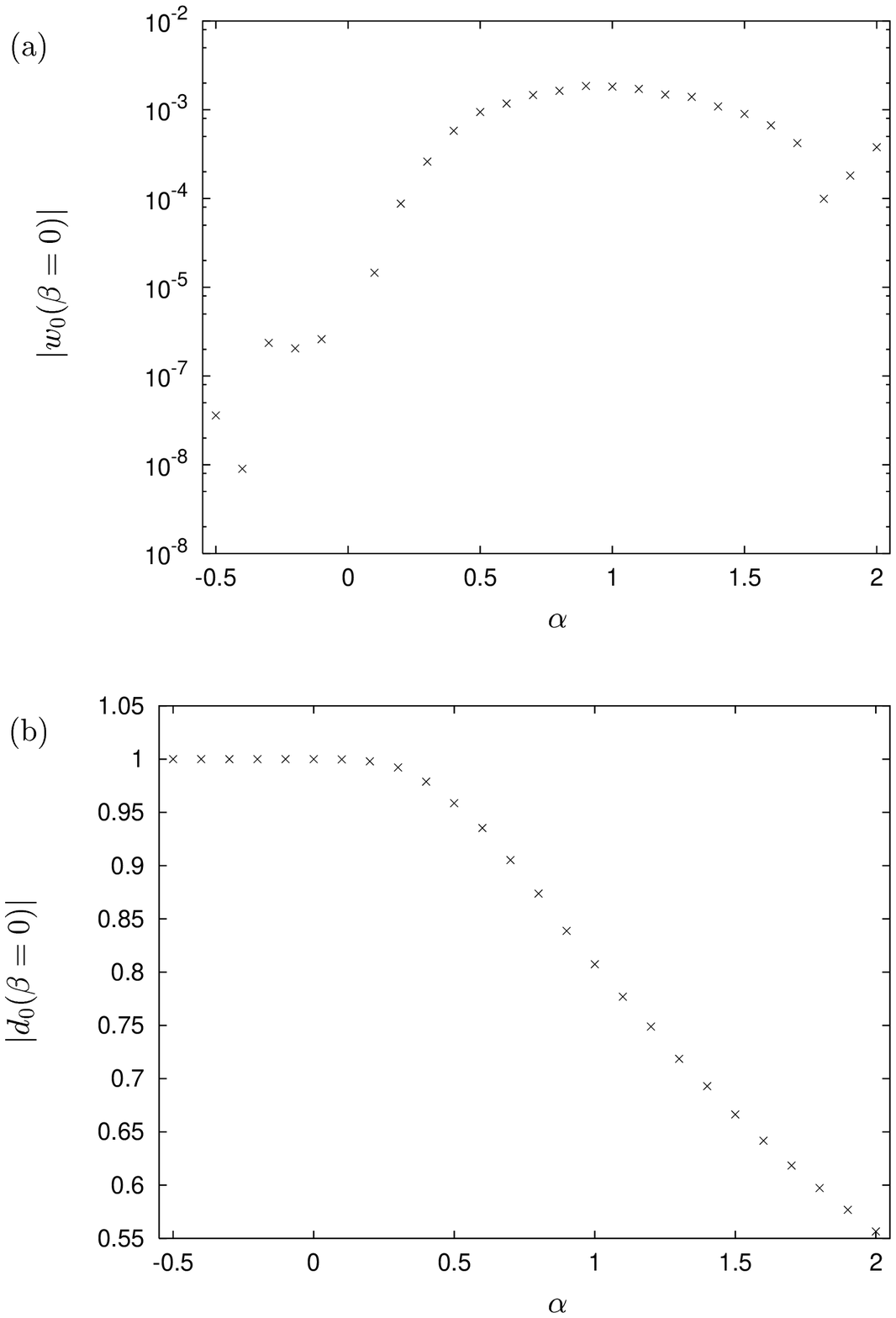}
\caption{
Test of the convergence of the harmonic inversion procedure:
Absolute value of the leading frequency $w_0$ for $\beta=0$ and the 
corresponding amplitudes $d_0$  obtained by harmonic inversion,
as a function of the parameter $\alpha$.
The closeness of these values to their theoretical values $w_0(\beta=0)=0$
and $d_0(\beta=0)=1$ are a measure for the reliability of the 
results for the diffusion constant obtained for the respective value of 
$\alpha$.}
\label{fig2}
\end{figure}

In our calculations, we observed that
the condition that for given $\alpha$
the expression (\ref{D-w0}) for the diffusion constant should become 
independent of $\beta$ for $\beta\to 0$ was well fulfilled for $\alpha<0$.
However, with increasing $\alpha>0$,
the results became numerically unstable if $\beta$ was too close to zero,
so that we had to extrapolate from the results for $\beta$ around $0.5$.
The convergence of the
frequencies and amplitudes in the harmonic inversion procedure itself
was also very good for $\alpha<0$, but convergence was increasingly
hard to obtain for $\alpha>0$.
A criterion for the reliability of the results for a given value
of $\alpha$ is the closeness of the leading frequency for $\beta=0$
to its theoretical value $w_0=0$, as well as the closeness of the
corresponding amplitude $d_0$ (see Eq.~(\ref{cex})) obtained by harmonic 
inversion to its theoretical value 1 (the multiplicity of the leading 
zero of the zeta function).
The absolute value of the leading frequency for $\beta=0$ and the
corresponding amplitude as a function of $\alpha$
are presented in Figure \ref{fig2}.
The absolute value of the leading frequency was smallest 
for $\alpha<0$ but stayed below
$|w_0|\approx 10^{-3}$ for all values of $\alpha$.
The corresponding amplitudes were nearly
exactly equal to their theoretical value 1 for $\alpha<0$,
but began to deviate from their theoretical value
for $\alpha\gtrsim 0.3$, indicating that the
harmonic inversion results became less well converged.
A possible reason for the deviations may be that the
density of frequencies becomes too large for the signal length chosen.
It can be expected that the accuracy of the 
results will be improved by including longer orbits.

In Ref.~\cite{Det97} it was proposed to truncate the periodic orbit sum
by introducing a stability cutoff and to use the stability as an ordering
parameter for the cycle expansion.
It is worth noting that the harmonic inversion procedure has the advantage 
that no discrete ordering parameter is needed at all.
However, a stability cutoff can also be used with the harmonic inversion 
technique because very unstable orbits give only a small contribution to 
the signal which can be neglected.

\section{Conclusion}
In conclusion, we have demonstrated how the diffusion constant
of one-dimen\-sional diffusive maps can be determined from a finite
set of periodic orbits using harmonic inversion techniques.
Starting from a zeta function whose leading zero possesses a known relation
to the diffusion constant, we have constructed a periodic orbit
signal which was analysed by harmonic inversion.
The diffusion constant could then be determined from the leading
frequency of the signal.
Since, in contrast to other methods such as cycle expansion,
harmonic inversion does not depend
on any special properties of the map, the general procedure should work
in the same way for all chaotic maps.
We have tested the method for a simple map 
whose periodic orbits can
easily be determined and for which results for the diffusion constant
obtained by cycle expansion techniques are available in the literature.
Our results are in excellent agreement with the ones from cycle expansion.

\section*{Acknowledgement}
We are grateful to R.~Grauer for turning our attention to the subject.
This work was supported by the Deutsche Forschungsgemeinschaft and
Deutscher Aka\-de\-mi\-scher Austauschdienst.


\end{document}